\documentclass[aps,preprint,amsmath,amssymb,showpacs,amsfonts]{revtex4}
\usepackage{epsfig}
\usepackage{graphicx}
\usepackage{dcolumn}
\usepackage{bm}
\usepackage{amsthm}
\usepackage{amsmath}
\usepackage{color}

\newcommand{\Lie}[1]{\mathcal{L}_{#1}\,}

\newcommand{\beq}{\begin{equation}}
\newcommand{\eeq}{\end{equation}}
\newcommand{\bea}{\begin{eqnarray}}
\newcommand{\eea}{\end{eqnarray}}

\begin{document}
\begin{flushright}
AEI-2012-042
\end{flushright}
\bigskip
\bigskip

\title{Local Entropy Current in Higher Curvature Gravity and\\ Rindler Hydrodynamics}
\author{Christopher Eling$^1$}
\author{Adiel Meyer$^2$}
\author{Yaron Oz$^2$}

\affiliation{$^1$ Max Planck Institute for Gravitational Physics, Albert Einstein Institute, Potsdam 14476, Germany}
\affiliation{$^2$ School of Physics and Astronomy, Tel Aviv University, Tel Aviv 69978, Israel}

\date{\today}
\begin{abstract}

In the hydrodynamic regime of field theories the entropy is upgraded to a local entropy current.
The entropy current is constructed phenomenologically order by order in the derivative expansion by requiring that its divergence is non-negative.
In the framework of the fluid/gravity correspondence, the entropy current of the fluid is mapped to a vector density associated with the event horizon of the dual geometry.
In this work we consider the local horizon entropy current for higher-curvature gravitational theories proposed in arXiv:1202.2469, whose flux
for stationary solutions is the Wald entropy. In non-stationary cases this definition contains ambiguities, associated with absence of a preferred timelike Killing vector. We argue that these ambiguities can be eliminated in general by choosing the vector that generates the subset of diffeomorphisms preserving a natural gauge condition on the bulk metric.  We study a dynamical, perturbed Rindler horizon in Einstein-Gauss-Bonnet gravity setting and compute the bulk dual solution to second order in fluid gradients. We show that the corresponding unambiguous entropy current at second order has a manifestly non-negative divergence.

\end{abstract}


\maketitle

\tableofcontents

\newpage

\section{Introduction}

According to the holographic principle \cite{'tHooft:1993gx,Susskind:1994vu}, quantum gravitational theories are equivalent to certain non-gravitational field theories living in one lower spatial dimension and defined on a boundary surface in the higher dimensional bulk spacetime. The only concrete realization of holography that we currently possess is based on the AdS/CFT correspondence between conformal field theories and their various deformations and quantum gravity (string theory) on
gravitational backgrounds with negative cosmological constant (for a review see \cite{Aharony:1999ti}). The mysterious nature of holography in general is a crucial aspect of the puzzle of quantum gravity. On a less fundamental level, holography and the AdS/CFT correspondence also offer new ways of investigating unresolved issues both in field theory and in gravitation.

One important example of this is the hydrodynamic regime of field theory \cite{Policastro:2001yc},
and the fluid-gravity correspondence \cite{Bhattacharyya:2008jc}, which originally followed as a special case of AdS/CFT. The hydrodynamics associated with a thermal state in the gauge theory is equivalent to the long wavelength, long time dynamics of black hole (brane) solutions in the bulk gravity theory. One can explicitly construct perturbed black hole solutions order by order as an expansion in derivatives of the fluid velocity and temperature and find that the subset of Einstein equations constraining data on the boundary surface are the Navier-Stokes equations.
The essential ingredients needed to relate fluids to gravity are the existence of a horizon
in the gravitational background that is related to a thermal equilibrium state in the field theory, and a derivative expansion around it \cite{Damour,Eling:2009pb,Eling:2009sj}.
Thus, one can define the relation between fluids and gravity
on more general backgrounds, for instance the Rindler geometry \cite{Bredberg:2011jq,Compere:2011dx,Compere:2012mt,Eling:2012ni}.

On the gravitational side of the duality, the correspondence has motivated new studies of black hole entropy in a dynamical setting, where the horizon surface evolves in time. In hydrodynamics the relevant quantity is a local entropy current. In a regime sufficiently close to equilibrium, the fluid-gravity mapping implies that the entropy current of the fluid flow can be constructed in terms of the event horizon geometry as the Bekenstein-Hawking \cite{Bekenstein:1973ur,Hawking:1974sw} area current \cite{Bhattacharyya:2008xc, Booth:2010kr}. The thermodynamical Second Law enforcing the positivity of the entropy current's divergence is equivalent to Hawking's area theorem in classical General Relativity. Further away from equilibrium it has been suggested that the correct hypersurfaces in the bulk on which to build the entropy current should be the quasi-locally defined apparent horizons (see for example, \cite{Figueras:2009iu,Booth:2011qy,Meyer:2011uka}), which also obey the Hawking theorem.

The main goal of this paper is to explore within the fluid setting a related issue in semi-classical gravity, which is the nature of dynamical horizon entropy in a higher curvature theory of gravity. In General Relativity, black hole thermodynamics allows one to clearly identify the entropy associated with equilibrium processes as proportional to the cross-sectional area of the event horizon. Outside the equilibrium setting, the key requirement for defining an entropy current phenomenologically is that it is consistent with the (generalized) Second Law. Thus, the link between entropy and area still seems to be robust due to Hawking's area theorem
the ambiguity is which horizon surface in the bulk is the appropriate holographic surface in general.

In higher curvature theories of gravity the situation becomes more complicated. Wald \cite{Wald:1993nt} studied quasi-stationary processes in a general diffeomorphism invariant theory of gravity and was able to derive a general formula for the entropy. The relevant quantity is the antisymmetric Noether potential $Q^{AB}$ associated with diffeomorphisms along a vector $\ell^A$. We will focus on the case where the gravitational Lagrangian depends only algebraically on the Riemann tensor $\mathcal{L}(g_{AB}, R_{ABCD})$. Here the potential has the form \cite{Jacobson:1993vj,LopesCardoso:1999cv}
\begin{align}
Q^{AB} = \sqrt{-g} \left(-2 \mathcal{L}^{ABCD} \nabla_C \ell_D + 4 \ell_D \nabla_C \mathcal{L}^{ABCD}\right)  \ , \label{Noethercurrent}
\end{align}
with $\mathcal{L}^{ABCD} =  \partial \mathcal{L}/\partial R_{ABCD}$\footnote{In the case where the Lagrangian depends only on the Riemann tensor algebraically, the current only involves first derivatives of $\ell^A$. For a general dependence on Riemann the current will involve higher order derivatives, which can be reduced to first derivatives using identity $\nabla_A \nabla_B \ell_C = R^{D}{}_{ABC} \ell_D$ which holds only if $\ell^A$ is a Killing vector.}. When the black hole solution is stationary, $\ell^A$ is the timelike Killing vector. The total entropy of the horizon is unambiguous and is proportional to an integral of $Q^{AB}$ over any horizon cross-section
\begin{align}
S_{wald} = \frac{1}{T} \int Q^{AB} d\Sigma_{AB} \ . \label{totalWald}
\end{align}
The bifurcation surface where $\ell^A$ vanishes and $\nabla_{[A} \ell_{B]} = \epsilon_{AB}$ is a convenient choice for actual computations \footnote{Note that for this reason the second term in (\ref{Noethercurrent}) does not contribute the Wald entropy of a stationary horizon}. In Einstein gravity the Wald formula reduces to the Bekenstein-Hawking area entropy, but in general the entropy will depend on the both the intrinsic and the extrinsic geometries of the horizon surface.

When the horizon is dynamical, the Wald formula is subject to ambiguities because there is no longer a preferred choice of the Killing vector and because one is free to add total divergence to the Lagrangian, symplectic potential, and (\ref{Noethercurrent}) itself \cite{Jacobson:1993vj,Iyer:1994ys}. 
In the paper we address the ambiguity in the definition of the vector $\ell^A$, and assume that the Noether potential has the standard form (\ref{Noethercurrent}) used by Wald.

Furthermore, there is no known analog of the area theorem in a general higher curvature theory, so the Second Law provides no guidance on how to appropriately define an entropy. In the context of the fluid-gravity correspondence, a definition for a local entropy current has been recently proposed \cite{Chapman:2012my}, which in a coordinate free form reads
\begin{align}
s^A = \frac{2\pi}{\kappa} Q^{AB} \ell_B \ . \label{entropydef}
\end{align}
This is to be evaluated on the horizon, so $\ell_B$ is the normal to the horizon and $\kappa$ is the surface gravity associated with this normal vector, defined in general as a measurement of its non-affinity, $\ell^B \nabla_B \ell^A = \kappa \ell^A$. Thus the current is effectively a flux of the Noether potential through the horizon surface. This expression contains an ambiguity in the definition of the vector $\ell^A$ off the horizon surface in the bulk. In particular the current $s^A$ depends on the derivative of $\ell^A$ with respect to the bulk radial coordinate. 
In \cite{Chapman:2012my} it was shown that this ambiguity could be eliminated in Einstein's gravity by requiring the vector field to satisfy a ``weak Killing condition" at the horizon
\begin{align}
\ell^A (\nabla_A \ell_B + \nabla_B \ell_A) = 0 \ . \label{weakKillingGen}
\end{align}
When this condition is imposed, the current in (\ref{entropydef}) reduces to the Bekenstein-Hawking area current. In higher curvature theories, there was apparently no general way to resolve the ambiguity. However, working to first order in fluid gradients, it was shown that the weak Killing condition (imposed at this order) again leads to a non-ambiguous current \cite{Chapman:2012my}. This current was constructed explicitly in the case of a charged black brane background in Einstein-Maxwell theory with a mixed gauge-gravitational Chern-Simons term and its divergence was shown to be non-negative \cite{Chapman:2012my}.

An important question is, whether the ambiguities in (\ref{entropydef}) can continue to be eliminated at higher orders in fluid gradients, where the theory is further away from the equilibrium state.
We propose that this can be naturally done by imposing a more general condition on $\ell^A$ (of which the weak Killing condition (\ref{weakKillingGen}) is just one). This is
\begin{align}
\Lie{\ell} g_{r A} = 0 \ ,    \label{Killing1}
\end{align}
where the coordinates are $X^A = (r,x^\mu)$ and $\Lie{\ell}$ is the Lie derivative in the direction of $\ell$. $x^\mu$ are the gauge theory coordinates and $r$ is the holographic radial coordinate. This choice follows from the standard gauge imposed on the bulk metric in the fluid-gravity correspondence $g_{rr}=0,~~g_{r\mu} \sim u_{\mu}$, where $u_{\mu}$ corresponds to the fluid velocity.
The construction of the bulk metric in a derivative expansion is carried out such that this gauge condition is preserved.
With this gauge choice, one has a map between the boundary and the horizon coordinates, and the hydrodynamics is realized in the $x^{\mu}$ space.
The conditions (\ref{Killing1}) that we propose determines $\ell^A$, such that it is preserves the gauge condition.
This is, in fact, a necessary condition if we want the entropy current constructed from the horizon data to be compatible with the hydrodynamics
determined by the boundary stress-energy tensor.
However, while this choice removes the ambiguity in the definition of $\ell^A$, we do not know whether it guarantees that the entropy current has
a non-negative divergence.

In this work we will investigate the nature of the entropy current in a particular setup. We will first construct the full second order solution to Einstein-Gauss-Bonnet gravity (in spacetime dimension greater than four) using the recently discovered fluid/Rindler correspondence \cite{Bredberg:2011jq,Compere:2011dx,Compere:2012mt,Eling:2012ni}. This correspondence is based on the fact that the Rindler wedge of flat Minkowski space acts as finite temperature thermal state analogous to the black brane in AdS. Furthermore, like the black brane, the Rindler acceleration horizon has a planar topology which allows as discussed above, to make the same controlled expansion in fluid gradients as in the fluid-gravity correspondence. Using this machinery, relativistic solutions to the vacuum Einstein equations have been constructed to second order \cite{Compere:2012mt,Eling:2012ni}. These bulk solutions are dual to a fluid system living on an arbitrary timelike surface $S_c$ of fixed radial coordinate $r=r_c$.

So far this is a holographic mapping between two classical theories, but it hints towards a full duality between some quantum field theory on $S_c$ and the interior region of the Rindler geometry. While the nature of holography in asymptotically flat spacetimes is a mystery, it is possible that some information can be gleaned from the non-standard properties of the dual fluid. In particular, the dual fluid thermodynamics is characterized by zero equilibrium energy density even though there is non-zero temperature. The corresponding viscous hydrodynamics of the system is perfectly well-defined, but has the intriguing property that only transport coefficients at second (and higher) order in the derivative expansion are affected by higher curvature terms in the gravitational theory. This implies the shear viscosity to entropy density ratio of the fluid is universal \cite{Chirco:2011ex}. Exploring the hydrodynamics of this fluid in more detail is of interest.

Following the procedure developed in \cite{Compere:2012mt,Eling:2012ni}, we compute the vacuum solution to second order. Since the Rindler background is flat, some simplifications occur and we are able to obtain the Gauss-Bonnet corrections to the metric and the fluid stress tensor. This extends the earlier results of \cite{Chirco:2011ex} in the non-relativistic limit to the fully relativistic fluid case.
In particular, we will find to first order in fluid gradients the extension of the horizon normal into the bulk,
\begin{align}
\ell^A \frac{\partial}{\partial X^A} = p^{-1} u^\mu \frac{\partial}{\partial x^\mu} \ , \label{ellvector}
\end{align}
and show that the entropy divergence at second order is non-negative.

The plan of this paper is as follows. In Section II we will review the correspondence between hydrodynamics and gravity and motivate the entropy current formula (\ref{entropydef}) in more detail. We discuss the nature of its possible ambiguities and why (\ref{Killing1}) is a natural condition to fix $\ell_A$. In Section III we construct the solution to vacuum Einstein-Gauss-Bonnet gravity up to second order in fluid gradients by perturbing around the Rindler background and find the holographic fluid stress tensor at second order. In Section IV we compute the entropy current using (\ref{entropydef}) and show that its divergence is non-negative. In the discussion, we examine the  implications of our results and possible extensions to a wider class of examples in the fluid-gravity paradigm and beyond. Finally, the appendices contain a more detailed discussion of ambiguities and specific calculations of the entropy current.

\section{The Entropy current}

We begin with a brief review of how the hydrodynamics of a fluid system can be encoded in a gravitational solution in one higher dimension. The key underlying concept in hydrodynamics is the notion of local thermodynamic equilibrium. The fluid is described by a finite set of macroscopic parameters, which are functions of space and time that vary slowly throughout the system, so that in the neighborhood of each point there is an approximate notion of thermodynamic equilibrium.  Thus, a relativistic fluid is characterized by a four-velocity $u^\mu$ and the thermodynamic variables, energy density $\rho$, pressure $p$, temperature $T$, and entropy density $s$. If the fluid also possesses additional conserved charges, one includes the corresponding charge densities and chemical potentials in its description. The various thermodynamic variables are related by the equation of state and by the standard equilibrium thermodynamical identities.

Hydrodynamics is an effective description valid at scales $L\gg \ell_{mfp}$, where $\ell_{mfp}$ is the mean free path of the system, determined by the temperature and the nature of the microscopic (field) theory. This means the gradients of all the fluid parameters must be small compared to the scale set by the mean free path and thus hydrodynamics is characterized by an expansion in derivatives of the fluid variables. At zeroth order, the fluid is in equilibrium and the entropy current $s u^\mu$ is conserved. Higher orders in derivatives correspond to viscous corrections, which are in general associated with dissipation and increase the entropy of the system.

On the gravity side, we consider a $(d+2)$ dimensional spacetime and denote the bulk coordinates by $X^A$, where the index $A$ runs from $(0..d+1)$. Typically, one decomposes $X^A = (r,x^\mu)$, where $r$ is the holographic radial coordinate and $x^\mu$  are the coordinates in the field theory/fluid. The index $\mu$ runs from $(0...d)$ and therefore $d$ is the number spatial dimensions of the fluid system. The metric ansatz corresponding to a fluid in global equilibrium is
\begin{align}
ds^2 = g_{AB} dX^A dX^B = k(r) u_\mu u_\nu dx^\mu dx^\nu - 2 u_\mu dx^\mu dr + f(r) P_{\mu \nu} dx^{\mu} dx^{\nu} \ . \label{ansatz}
\end{align}
Here, $u^\mu u_\mu = -1$ and $u_{\mu}$ can be thought of as the velocity of a boost. $P_{\mu \nu} = h_{\mu \nu} + u_\mu u_\nu$ is the projector orthogonal to $u^\mu$ and $h_{\mu \nu}$ the metric (possibly curved) on which the fluid system lives. The functions $f(r)$ and $k(r)$ are determined by the field equations. There is an event horizon located at radius $r_h$ such that $k(r=r_h)=0$. At this location, the Eddington-Finkelstein like coordinates chosen for (\ref{ansatz}) are regular. To see that this metric corresponds to a holographic fluid, one can compute the Brown-York stress tensor for a surface of constant $r$ and show it has the form of a perfect fluid \cite{Bredberg:2010ky}
\begin{align}
T_{\mu \nu} = \rho u_\mu u_\nu + p P_{\mu \nu} \ ,
\end{align}
with Hawking temperature
\begin{align}
T = -\frac{\partial_r k}{4\pi}|_{r=r_h} \ .
\end{align}
Note, that in order to describe charged fluids, extra gauge fields are needed in the ansatz.

The extension to an arbitrary fluid state is straightforward: one promotes $u^\mu(x^\mu)$, $k(r,x^\mu)$ and $f(r,x^\mu)$. The metic ansatz is no longer an exact solution to the field equations, but one can work order by order in an expansion in derivatives of these variables as done in \cite{Bhattacharyya:2008jc}. The details of this construction for the Rindler metric will be described in the next section. Here we note that the perturbed metric solution implies the event horizon is dynamical and its location varies in time and space, $r_h(x^\mu)$. The horizon location is determined by solving the equation for a null hypersurface
\begin{align}
g^{AB} \partial_A (r-r_h(x^\mu) \partial_B (r-r_h(x^\mu)) = 0 \ ,
\end{align}
order by order in the derivative expansion. The horizon normal vector $\ell^A$ then follows from $\ell^A = g^{AB}\partial_B (r-r_h(x^\mu))$.

When working with the the entropy current (\ref{entropydef}) it will be useful consider a coordinate gauge adapted to the horizon $X^A = (\bar{r}, x^\mu)$, where the horizon is always located at zero radius $\bar{r}=0$, i.e.
\begin{align}
\bar{r} = r-r_h(x) \ .
\end{align}
In these coordinates
\begin{align}
\ell^{\bar{r}} = g^{\bar{r} \bar{r}}|_{\bar{r} = 0} = 0 \nonumber \\
\ell^\mu = g^{\bar{r} \mu}|_{\bar{r} = 0} \ ,
\end{align}
and the entropy current (\ref{entropydef}) reduces to
\begin{align}
s^A = (0, s^\mu) \ ,
\end{align}
where
\begin{align}
s^\mu = \frac{2\pi}{\kappa} Q^{\mu \bar{r}} = \frac{2\pi}{\kappa} (-2 \mathcal{L}^{\mu \bar{r} \nu \bar{r}}(\nabla_\nu \ell_{\bar{r}}
   -\nabla_{\bar{r}} \ell_\nu) + 4 \nabla_\nu \mathcal{L}^{\mu \bar{r}  \nu \bar{r} }) \label{entropycurrent} \ .
\end{align}

In the case of Einstein's gravity, where $\mathcal{L} = \sqrt{-g} R$ (we use units where $16\pi G = 1$),
\begin{align}
\mathcal{L}^{ABCD} = \frac{1}{2} \left(g^{AC} g^{BD} - g^{AD} g^{BC}\right) \ ,
\end{align}
and the second covariant derivative term in (\ref{entropycurrent}) vanishes identically. In equilibrium one can use the Killing equation $\nabla_A \ell_B = -\nabla_B \ell_A$ to set $\nabla_{\bar{r}} \ell_\mu = - \nabla_\mu \ell_{\bar{r}}$. As a result, (\ref{entropycurrent}) reduces to the Bekenstein-Hawking area current
\begin{align}
s^\mu_{GR} = 4\pi \sqrt{-g} \ell^\mu \ ,
\end{align}
which for the metric ansatz above (\ref{ansatz}) reduces to $s^\mu = 4\pi f(0)^{3/2} u^\mu$.

In the dynamical case, there is no longer a Killing vector, but it is possible to use the freedom in the radial derivative of $\ell_A$ (now thought of as a generalization of the Killing vector in the bulk which becomes the null normal when evaluated on the horizon) to impose the ``weak Killing condition" \cite{Chapman:2012my}
\begin{align}
\ell^A (\nabla_A \ell_B + \nabla_B \ell_A) = 0 \ .
\end{align}
Since the $\mu$ component of the equation turns out to be an identity, this amounts to one condition
\begin{align}
\ell^B (\nabla_{\bar{r}} \ell_B + \nabla_B \ell_{\bar{r}}) = \frac{1}{2} \nabla_{\bar{r}} (\ell_B \ell^B) + \kappa \ell_{\bar{r}} = 0 \label{weakKilling0} \ ,
\end{align}
When evaluated on the horizon this condition relates the surface gravity at higher orders to the radial derivative of the norm of $\ell^A$. With this condition imposed, $\ell^\mu \nabla_{\bar{r}} \ell_\mu = - \ell^\mu \nabla_\mu \ell_{\bar{r}}$,  we get that definition (\ref{entropycurrent}) again reduces to the Bekenstein-Hawking current. Note, that at this stage the derivative expansion has not played a role. As long as weak Killing condition can be enforced in a general dynamical situation, (\ref{entropydef}) always reduces to the area current.

In a higher curvature theory, one must restrict to the fluid-gravity setting and analyze the nature of the ambiguities order by order in the derivative expansion. This was done in \cite{Chapman:2012my} for a generic higher curvature theory to first order in fluid derivatives. The result is that the weak Killing condition (at first order) is again sufficient to eliminate the ambiguity in the definition of the entropy current. In Appendix A, we continue this type of analysis to second order in gradients. Here one needs a new set of constraints on the components of the radial derivative of $\ell^A$ in addition to the weak Killing condition at second order. It turns out that these conditions ultimately follow from the positivity of the entropy current divergence.

Instead, here we propose a natural, geometrical way to eliminate all the ambiguities \textit{a priori} and in general is to simply impose the stronger condition in (\ref{Killing1}). This set of $(d+2)$ equations fixes the $(d+2)$ components in $\ell^A$, from this one can determine the radial derivatives that appear in the formula (\ref{entropydef}).  With (\ref{Killing1}) imposed, $\ell^A$ is the vector generating coordinate transformations that preserve the bulk gauge condition typically employed in the fluid-gravity correspondence \cite{Bhattacharyya:2008ji},
\begin{align}
g_{r r}=0; \quad g_{r \mu} = -u_\mu \ .
\end{align}
Geometrically, in this gauge lines of constant $x^\mu$ are null geodesics and $r$ is the affine parameter along these geodesics. This choice of gauge is associated with a trivial mapping of points on the arbitrary boundary hypersurface to the event horizon along the ingoing null geodesics, i.e. the gauge theory coordinates $x^\mu$ are also coordinates on the event horizon hypersurface and $r_h(x^\mu)$. Note that the form of this gauge is the same in both the $r$ and $\bar{r}$ coordinate systems, so that ultimately either choice will suffice.  In the following sections, we will compute the perturbed Rindler solution to second order and demonstrate that the now unambiguous entropy current (\ref{entropydef}) has a non-negative divergence, consistent with the generalized Second Law.

\section{Perturbed Rindler solution in vacuum Einstein-Gauss-Bonnet gravity}

Following (\ref{ansatz}) the metric for the flat Rindler wedge can be expressed in the following form \cite{Compere:2012mt,Eling:2012ni}
\begin{align}
ds^2 = g_{AB} dx^A dx^B = -(1+p^2(r-r_c)) u_\mu u_\nu dx^\mu dx^\nu - 2 p u_\mu dx^\mu dr + P_{\mu \nu} dx^\mu dx^\nu  \ . \  \label{zerothmetric}
\end{align}
The holographic fluid system lives on the surface $r=r_c$ and has a pressure
\beq
p = \frac{1}{\sqrt{r_c - r_h}} \ .
\eeq
Here $P_{\mu \nu} = \eta_{\mu \nu} + u_\mu u_\nu$, so the fluid is on the ordinary flat Minkowski metric. $r_h>0$ is the location of the horizon. Note that in this section we will use the standard radial coordinate $r$ (that is in some sense adapted to the boundary) instead of the horizon adapted $\bar{r}$. Evaluating the holographic (Brown-York) stress tensor at $r=r_c$ yields
\beq
T_{\mu \nu} = p P_{\mu \nu} \ ,
\eeq
which indicates the Rindler metric is dual to a fluid with zero equilibrium energy density $\rho = 0$.

A flat metric is not only a vacuum solution to Einstein equation, but also to any standard higher curvature theory, where the action is constructed from higher powers of the Riemann tensor and its derivatives. Here we will consider the Einstein-Gauss-Bonnet theory, which is given by the action
\beq \mathcal{L}_{GB} = \int d^{d+2} x \sqrt{-g} \left[R + \alpha \left (R^2 - 4 R_{CD} R^{CD} + R_{CDEF} R^{CDEF}\right) \right] \ , \label{GBaction}
\eeq
where $\alpha$ is the Gauss-Bonnet coupling constant. We consider $d \geq 3$ since for $d<3$ the Gauss-Bonnet term is topological and does not affect the field equations. There are two reasons to consider the Gauss-Bonnet term. First, Einstein-Gauss-Bonnet gravity is notable because even though the action is higher order in the curvature, for the unique combination of curvature invariants in the second term of (\ref{GBaction}), the field equations remain second order in derivatives of the metric. Second, such a term often arises in the low energy limit of string theories.

The field equations are given by
\beq
G_{AB} + 2 \alpha H_{AB} = 0 \ , \label{EGBfieldeqn}
\eeq
where the tensor $H_{AB}$ is defined as
\begin{align}
H_{AB} &= R R_{AB} - 2 R_{AC} R^C_B - 2 R^{CD} R_{ACBD} + R_A{}^{CDE} R_{BCDE} \nonumber \\ & - \frac{1}{4} g_{AB} \left(R^2 - 4 R_{CD} R^{CD} + R_{CDEF} R^{CDEF}\right)
\ .  \label{Lovelock}
\end{align}
To put field equations in a more compact form, we take the trace of (\ref{EGBfieldeqn}), which leads to the following on-shell condition:
\beq
R = \frac{4\alpha}{d} H_C^C \ .
\eeq
Substituting back into the field equations, we find
\beq
Y_{AB} \equiv R_{AB} + 2\alpha X_{AB} = 0 \label{eom} \ ,
\eeq
where
\beq X_{AB} = H_{AB} - \frac{1}{d} g_{AB} H_C^C \ . \eeq

To model the Rindler fluid in local equilibrium, one allows $u^\mu(x^\mu)$ and $p(x^\mu)$, but leaves $r=r_c$ and the induced flat metric on it fixed.  The metric is no longer a solution to the field equations, but one can expand and work order by order in derivatives of the fields $u^\mu$ and $p$. The metric (\ref{zerothmetric}) is a solution at zeroth order, i.e. $Y_{AB} = 0 + O(\lambda)$, where $\lambda$ is a parameter that counts the derivatives of $u^\mu$ and $p$. The strategy for solving the equations is as follows. One introduces a metric correction at first order
\beq
 g = g^{(0)} + \delta g^{(1)} \ .
\eeq
The corrected metric at first order induces a $\delta Y^{(1)}_{AB}$ at the same order, which means it involves only radial derivatives. We want to solve for the metric $\delta g^{(1)}$ so that
\beq \delta Y^{(1)}_{AB} + \hat{Y}^{(1)}_{AB} = 0  \ ,\eeq
where $\hat{R}^{(1)}_{AB}$ comes from the zeroth order metric. This method can be generalized to solve for the metrics at higher order in $\lambda$. If we have a solution to $(n-1)$ order $g^{(n-1)}$, then one introduces a correction $\delta g^{(n)}$ so that
\beq \delta Y^{(n)}_{AB} + \hat{Y}^{(n)}_{AB} = 0 \ .  \label{fieldeqn} \eeq

Expanding out, we find
\beq
\delta R^{(n)}_{AB} + \hat{R}^{(n)}_{AB} + 2 \alpha(\delta X^{(n)}_{AB} + \hat{X}^{(n)}_{AB}) = 0 \ .
\eeq
There is no change to the $H_{AB}$ tensor (\ref{Lovelock}) at the same order $n$ since the curvature of the Rindler background is zero and any term in the variation would contain some factor of curvature at zero order. Thus, $\delta X^{(n)}_{AB}=0$. At first viscous order ($n=1$) the results found previously for GR hold, because $\hat{H}^{(1)}_{AB}=0$. This follows just from the fact that the lowest order part of the Riemann tensor is first order for the Rindler solution and $H_{AB}$ tensor is quadratic the Riemann tensor and its contraction. In \cite{Chirco:2011ex} it was shown that for any higher curvature theory the first order corrections vanish and the viscous hydrodynamics is the same as in GR. The result for the dual metric to first order is \cite{Compere:2012mt,Eling:2012ni}
\begin{align}
ds^2 &= -(1+p^2(r-r_c)) u_\mu u_\nu dx^\mu dx^\nu - 2 p u_\mu dx^\mu dr + P_{\mu \nu} dx^\mu dx^\nu \nonumber \\
&+ 2 p (r-r_c) D(\ln p) u_\mu u_\nu  dx^\mu dx^\nu - 4 p (r-r_c)  u_{(\mu} P^{\lambda}_{\nu)} \partial_\lambda \ln p  dx^\mu dx^\nu  \ . \label{metric1}
\end{align}

One can show the general solution consistent with the boundary conditions takes the form
\begin{align}
P_\mu^\lambda P_\nu^\sigma \delta g^{(n)}_{\lambda \sigma} &= -2 p^2 \int^{r_c}_{r} \frac{1}{\Phi} dr' \int^{r'}_{r_c-\frac{1}{p^2}} P_\mu^\lambda P_\nu^\sigma \hat{Y}^{(n)}_{\lambda \sigma} dr'' \\
u^\lambda P^\sigma_\mu \delta g^{(n)}_{\lambda \sigma} &= (1/2) (1-r/r_c) V^{(n)}_\mu(x) - 2 p \int^{r_c}_{r} dr' \int^{r_c}_{r'} dr'' P^\lambda_\mu \hat{Y}^{(n)}_{r \lambda} \\
u^\lambda u^\sigma \delta g^{(n)}_{\lambda \sigma} &= (1-r/r_c) A^{(n)}(x) + p \int^{r_c}_{r} dr' \int^{r_c}_{r'} dr'' \left(p P^{\lambda \sigma} \hat{Y}^{(n)}_{\lambda \sigma} - p^{-1} \Phi \hat{Y}^{(n)}_{rr} - 2 \hat{Y}^{(n)}_{r \lambda} u^\lambda \right) \ , \label{solution}
\end{align}
where $\hat{Y}^{(n)}_{AB} = \hat{R}^{(n)}_{AB} + 2\alpha \hat{X}^{(n)}_{AB}$. From here on we consider the second order viscous calculation, which is where the higher curvature corrections first appear. We only need to find $\hat{H}_{AB}$ at 2nd order, which amounts to just
\begin{align}
\hat{H}^{(2)}_{AB} =  \hat{R}^{(1)}_A{}^{CDE} \hat{R}^{(1)}_{BCDE} -  \frac{1}{4} g^{(0)}_{AB} \hat{R}^{(1)}_{CDEF} \hat{R}^{(1) ~ CDEF} \ .
\end{align}
We find
\begin{align}
\hat{H}^{(2)}_{rr} &= 0 \\
\hat{H}^{(2)}_{r\mu} &= 0 \\
\hat{H}^{(2)}_{\mu \nu} &= -\frac{3}{2} p^2 a^\beta ( u_{(\nu}\partial_{\mu)} u_\beta -  u_{(\nu} \partial_{|\beta|} u_{\mu)}) + \frac{3}{2} p^2 \partial_{(\mu} u^\lambda \partial_{|\lambda|} u_{\nu)}  \nonumber \\
& - \frac{3}{4} p^2 P^{\beta \lambda} \partial_\beta u_\mu \partial_\lambda u_\nu - \frac{3}{4} p^2 \partial_\mu u^\beta \partial_\nu u_\beta +  \frac{3}{4} p^2 u_\mu u_\nu a^\lambda \partial_\lambda \text{ln} p  + \frac{3}{2} P_{\mu \nu} \Omega_{\alpha
\beta} \Omega^{\alpha \beta} \ .
\end{align}
Since only the $(\mu \nu)$ components of $H^{(2)}_{AB}$ are non-zero, for the solution (\ref{solution}),
\begin{align}
\hat{X}^{(2)}_{rr} &= 0 \\
\hat{X}^{(2)}_{r\mu} &= \frac{p}{d} u_\mu \hat{H}^{(2)}{}_C^C \\
\hat{X}^{(2)}_{\mu \nu} &= \hat{H}^{(2)}_{\mu \nu} - \frac{1}{d}(-\Phi u_\mu u_\nu + P_{\mu \nu}) \hat{H}^{(2)}{}_C^C \ .
\end{align}
Hence we only need the trace $\hat{H}^{(2)}{}_C^C = P^{\mu \nu} H^{(2)}_{\mu \nu}$ and $P^\lambda_\mu P^\sigma_\nu \hat{H}^{(2)}_{\lambda \sigma}$. Using the variables $\mathcal{K}_{\mu \nu} = P^\lambda_\mu P^\sigma_\nu \partial_{(\lambda} u_{\sigma)}$, $\Omega_{\mu \nu} = P_\mu^\lambda P_\nu^\sigma\partial_{[\lambda}u_{\sigma]}$, and $D^\perp_\mu = P^\nu_\mu \partial_\nu$
we find
\begin{align}
\hat{H}^{(2)}{}_C^C = + \frac{3}{2} p^2 (d-2) \Omega_{\alpha \beta} \Omega^{\alpha \beta} \ ,
\end{align}
and
\begin{align}
P^\lambda_\mu P^\sigma_\nu \hat{H}^{(2)}_{\lambda \sigma} &= + 3 p^2 \Omega_{\mu}{}^\lambda \Omega_{\lambda \nu} +  \frac{3}{2} p^2 P_{\mu \nu} \Omega_{\alpha \beta} \Omega^{\alpha \beta} \ .
\end{align}

Putting all this together we find using (\ref{solution})
\begin{align}
P_\mu^\lambda P_\nu^\sigma \delta g^{(2)}_{\lambda \sigma} &= \left(GR solution\right) + 6 \alpha p^2 (r-r_c) \left(\Omega_{\mu}{}^\lambda \Omega_{\lambda \nu} + \frac{1}{d} P_{\mu \nu} \Omega_{\alpha \beta} \Omega^{\alpha \beta}\right) \label{g1}\\
u^\lambda P^\sigma_\mu \delta g^{(2)}_{\lambda \sigma} &= (1/2) (1-r/r_c) V^{(2)}_{\mu}(x) + \left(GR solution\right) \label{g2}\\
u^\lambda u^\sigma \delta g^{(2)}_{\lambda \sigma} &= (1-r/r_c) A^{(2)}(x) + \left(GR solution\right)  + 3 \alpha p^4 (r-r_c)^2  \frac{(d-2)}{d} \Omega_{\alpha \beta} \Omega^{\alpha \beta}  \ . \label{g3}
\end{align}
To keep the equations compact we will not write the explicit form of the GR solution computed in \cite{Compere:2012mt,Eling:2012ni}. The free functions $A(x)$ and $V_{\mu}(x)$ are still arbitrary and may still depend on the Gauss-Bonnet coupling constant. The functions are fixed by imposing gauge conditions on the stress tensor.  In the Gauss-Bonnet theory, the form of the holographic stress tensor is modified to \cite{Davis:2002gn}
\beq
T_{\mu \nu} =  2 (K \gamma_{\mu \nu} - K_{\mu \nu}) + 4 \alpha (J \gamma_{\mu \nu} - 3 J_{\mu \nu}) \ ,  \label{stresstensor}
\eeq
where\footnote{Here an additional term is zero because the induced metric on $r=r_c$ is flat}
\beq
J_{\mu \nu} = \frac{1}{3} (2 K K_{\mu \sigma} K^\sigma_\nu + K_{\sigma \lambda} K^{\sigma \lambda} K_{\mu \nu} - 2 K_{\mu \sigma} K^{\sigma \lambda} K_{\lambda \nu} - K^2 K_{\mu \nu}). \label{Jstress}
\eeq

The first step to calculate the new stress tensor is consider the extrinsic curvature at second order induced by the second order metric. The result is
\begin{align}
\delta K^{(2)}_{\mu \nu} = \frac{1}{2p} \partial_r g^{(2)}_{\mu \nu}|_{r=r_c}  \ . \label{K2nd}
\end{align}
As a result the pure Brown-York part of the holographic stress tensor in linear in extrinsic curvature has the form
\begin{align}
\delta T^{(2)~ BY}_{\mu \nu} = \frac{1}{p} A^{(2)} P_{\mu \nu} + \frac{1}{p} u_{(\mu} V^{(2)}_{\nu)} + \left(GR solution\right) - 12 p \alpha \left(\Omega_{\mu}{}^\lambda \Omega_{\lambda \nu} + \frac{1}{d} P_{\mu \nu} \Omega_{\alpha \beta} \Omega^{\alpha \beta}\right) \ .
\end{align}
Next, we must consider the variation $\delta J^{(2)}_{\mu \nu}$. From (\ref{Jstress}) this involves the second order extrinsic curvature in (\ref{K2nd}) times two zeroth order $K^{(0)}_{\mu \nu}$. Due to the simple form
\begin{align}
K^{(0)}_{\mu \nu} = -\frac{p}{2} u_\mu u_\nu \ ,
\end{align}
one can show that $\delta J_{\mu \nu}$ vanishes identically and there is no contribution from the explicit $\alpha$ part of the holographic stress tensor.

In the explicit $\alpha$ contribution to the stress-tensor, we have to calculate $\hat{J}^{(2)}_{\mu \nu}$, which involves two first order parts of the extrinsic curvature times one zeroth order part. Using
\begin{align}
\hat{K}^{(1)}_{\mu \nu} = \mathcal{K}_{\mu \nu} \ ,
\end{align}
we find
\begin{align}
\hat{T}^{(2)~ J}_{\mu \nu} = 4 \alpha \left(\frac{1}{2} p P_{\mu \nu} \mathcal{K}_{\alpha \beta} \mathcal{K}^{\alpha \beta} - p \mathcal{K}_{\mu \lambda} \mathcal{K}^{\sigma}_{\mu}\right) \ .
\end{align}
Putting this all together, we arrive at
\begin{align}
T^{(2)}_{\mu \nu} &= \frac{1}{r_cp} A^{(2)} P_{\mu \nu} + \frac{1}{r_cp} u_{(\mu} V^{(2)}_{\nu)} + \left(GR terms\right) - 12 p \alpha \left(\Omega_{\mu}{}^\lambda \Omega_{\lambda \nu} + \frac{1}{d} P_{\mu \nu} \Omega_{\alpha \beta} \Omega^{\alpha \beta}\right) + \nonumber \\ & + 4 \alpha \left(\frac{1}{2} p P_{\mu \nu} \mathcal{K}_{\alpha \beta} \mathcal{K}^{\alpha \beta} - p \mathcal{K}_{\mu \lambda} \mathcal{K}^{\sigma}_{\mu}\right) \ .
\end{align}

The first gauge condition we require is that $T^{(2)}_{\mu \nu} u^\mu P^\nu_\lambda = 0$, which fixes $V^{(2)}_{\nu}$\footnote{Note that in \cite{Chapman:2012my} the authors also used the freedom in the function $V^{(n)}_{\mu}$ so that there are no higher order viscous corrections to the null normal, that is $\ell^A = (0, p^{-1} u^\mu)$ to all orders. The disadvantage of this frame choice in our case is that stress tensor will no longer satisfy $T^{(n)}_{\mu \nu} u^\mu P^\nu_\sigma = 0$. Hence we will remain in the standard Landau-like frame in the following.}. In this case there are no $\alpha$ corrections and as result (see Eqn. (\ref{g2})) $u^\lambda P^\sigma_\mu \delta g^{(2)}_{\lambda \sigma}$ is the same as in GR. The second condition is to cancel all terms proportional to $P_{\mu \nu}$ so that the pressure is unchanged at higher viscous orders. This fixes the $A^{(2)}$ in $u^\lambda u^\sigma \delta g^{(2)}_{\lambda \sigma}$ to be
\begin{align}
A^{(2)} = \left(GR result\right) +  \frac{12r_cp^2}{d} \alpha \Omega_{\alpha \beta} \Omega^{\alpha \beta} - 4 r_c p^2 \alpha \mathcal{K}_{\alpha \beta} \mathcal{K}^{\alpha \beta} \ .
\end{align}
The remaining contributions to the stress tensor are
\begin{align}
T^{(2)}_{\mu \nu} = \left(GR terms\right) - 12 \alpha p \Omega_{\mu}{}^\lambda \Omega_{\lambda \nu} - 4 \alpha p \mathcal{K}_{\mu \lambda} \mathcal{K}^{\lambda}_{\nu}.
\end{align}
Comparing to the general form of the fluid stress tensor \cite{Compere:2011dx}
\begin{align}
T_{\mu \nu} &= \rho u_\mu u_\nu + p P_{\mu \nu} - 2 \eta \mathcal{K}_{\mu \nu} \nonumber \\ & \quad + c_1 \mathcal{K}_\mu^\lambda \mathcal{K}_{\lambda \nu} + c_2 \mathcal{K}_{(\mu}^\lambda \Omega_{|\lambda|\nu)} + c_3 \Omega_\mu^{\,\,\,\lambda}\Omega_{\lambda \nu} +
c_4 P_\mu^\lambda P_\nu^\sigma D_\lambda D_\sigma \ln p \nonumber\\
& \quad  + c_5 \mathcal{K}_{\mu \nu}\,D\ln p + c_6 D^\perp_\mu \ln p \,D^\perp_\nu \ln p \ , \label{generalstress}
\end{align}
we see that the Gauss-Bonnet corrections only modify the $c_1$ and $c_3$ transport coefficients at second order to
\begin{align}
c_1 &= -\frac{2}{p}(1+2\alpha p^2) \nonumber\\
c_3 &= -\frac{4}{p}(1+3 \alpha p^2) \ , \label{GBcoeff}
\end{align}
with the rest of the coefficients $\eta$, $c_2$,$c_4$,$c_5$,$c_6$ the same as in GR. These results for the transport coefficients are consistent with \cite{Chirco:2011ex}, where the transport coefficients $c_{1..4}$ were read off from the non-relativistic solution.

\section{Rindler fluid entropy in Einstein-Gauss-Bonnet}

We now want to calculate the entropy current in the Einstein-Gauss-Bonnet case using the entropy current prescription in Section II.  We can express (\ref{entropycurrent}) as
\begin{align}
s^\mu_{EGB} = \frac{2\pi}{\kappa} \sqrt{-g} \left(-2 \mathcal{L}^{\mu \bar{r} \nu \bar{r}}_{tot} A_{\nu}\right).   \label{entropycurrentGB}
\end{align}
Here we have defined
\begin{align}
A_\mu \equiv \nabla_\mu \ell_{\bar{r}} - \nabla_{\bar{r}} \ell_\mu \ .
\end{align}
Note that in the case of a Einstein-Gauss-Bonnet (and its Lovelock generalizations \cite{Lovelock:1971yv}), the covariant derivative of $\mathcal{L}^{ABCD}$ on any index must vanish identically. In the $(\bar{r},x^\mu)$ coordinates
\begin{align}
\mathcal{L}^{\mu \bar{r} \nu \bar{r}}_{tot} = -\frac{1}{2} \ell^\mu \ell^\nu + \mathcal{L}^{\mu \bar{r} \nu \bar{r}}_{GB} \ , \label{Ltotal}
\end{align}
and
\begin{align}
\mathcal{L}^{\mu \bar{r} \nu \bar{r}}_{GB} = 2 \alpha\left(R^{\mu \bar{r} \nu \bar{r}} - R^{\bar{r} \bar{r}} g^{\mu \nu} + R^{\mu \bar{r}} \ell^\nu + R^{\nu \bar{r}} \ell^\mu - \frac{1}{2} R \ell^\mu \ell^\nu \right) \ . \label{LGB}
\end{align}
Here the expressions are to be always evaluated at the horizon $\bar{r}=0$.

Using the first order solution (\ref{metric1}), we find that $\Lie{\ell} g_{\mu A} = 0$ implies that $\ell^A$ takes the form in (\ref{ellvector}), just an extension of the null normal to first order off the horizon surface into the bulk. In terms of the variables in (\ref{entropycurrentGB}) we now have that
\begin{align}
u^\mu A^{(1)}_\mu &= 2 p \kappa^{(1)} \nonumber \\
\kappa^{(1)} &= p^{-1} D \ln p \nonumber \\
P^\lambda_\mu A^{(1)}_\lambda &= - 2 D^{\perp}_\mu \ln p \ , \label{firstdata}
\end{align}
where the superscript in parenthesis denotes the number of gradients in the expression.

At zeroth and first order, the Gauss-Bonnet current should also match the GR result since the first order solutions are the same, as we discussed above. Let's check this using (\ref{entropycurrentGB}). At zeroth order, we immediately get the the GR entropy current $s^\mu_{GB} = 4\pi u^\mu$ as expected.  In the first order solution (\ref{metric1}) it was found previously that $\ell^\mu_{(1)} = 0$ and $\sqrt{-g^{(1)}} = 0$.  Using
\begin{align}
\kappa^{(0)} &= \frac{1}{2} \nonumber\\
A^{(0)}_{\mu} &= - p u_\mu \ , \label{zerothdata}
\end{align}
and (\ref{firstdata}), at first order the expression is
\begin{align}
s^{\mu (1)}_{EGB} = 8\pi p^2 \mathcal{L}^{\mu \bar{r} \nu \bar{r}(1)}_{GB} u_\nu \ . \label{firstorderGB}
\end{align}
Then since
\begin{align}
R^{AB}_{(1)} = 0; \quad R_{(1)} = 0 \ ,
\end{align}
on-shell, many terms in $\mathcal{L}^{\mu \bar{r} \nu \bar{r}}$ in do not contribute at this order. The remaining term has the form
\begin{align}
\mathcal{L}^{\mu \bar{r} \nu \bar{r}(1)}_{GB} = 2 \alpha R^{\mu \bar{r} \nu \bar{r}(1)} = \frac{\alpha}{p} \mathcal{K}^{\mu \nu}, \label{firstorderL}
\end{align}
where we used the fact that
\begin{align}
R^{\mu \bar{r} \nu \bar{r} (1)} = R^{\mu r \nu r (1)}
\end{align}
with respect to components in the standard $r$ coordinate. Thus $s^{\mu (1)}_{EGB} = 0$ as expected since $\mathcal{K}^{\mu \nu} u_\nu = 0$.

Now we want to calculate the second order part of the current. It turns out that the first order data in (\ref{firstdata}) is sufficient to elminate most the ambiguities. We just need the additional second order condition , which follows from (\ref{Killing1}),
\begin{align}
u^\mu A^{(2)}_\mu = 2 p \kappa^{(2)} \ .
\end{align}
Then using (\ref{entropycurrentGB}), (\ref{Ltotal}) and (\ref{zerothdata}), the full current at second order takes the form
\begin{align}
s^{\mu (2)}_{EGB} &=  4\pi p^{-1} \sqrt{-g^{(2)}} u^\mu + 4\pi p\ell^\mu_{(2)} - 4\pi p u^\mu (\ell^\lambda_{(2)} u_\lambda) \nonumber \\ & + 8\pi p^2 \mathcal{L}^{\mu \bar{r} \nu \bar{r}(2)}_{GB} u_\nu +16\pi \alpha \mathcal{K}^{\mu \lambda}  D^{\perp}_\mu \ln p \ . \label{secondentropy}
\end{align}
The first line has no explicit $\alpha$ corrections, but there could in principle be Gauss-Bonnet corrections to the normal and metric determinant.

First consider the metric determinant. In our standard $r$ gauge for the metric, we have
\begin{align}
g_{rr} &= 0 \\
g_{r\mu} &= -p u_\mu \\
g_{\mu \nu} &= P_{\mu \nu} + corrections \ .
\end{align}
Thus, one can show
\begin{align}
\det g  = p \det (P^\lambda_\mu P^\sigma_\nu \delta g_{\lambda \sigma}) \ .
\end{align}
To second order, the transverse part of $g_{\mu \nu}$ is the identity matrix plus corrections because $P^\lambda_\mu P^\sigma_\nu g^{(1)}_{\lambda \sigma} =0$ (see (\ref{metric1})). Using the formula $\det(I+A) \approx 1 + Tr(A)$ gives
\begin{align}
\det g = p (1 + P^{\mu \nu} \delta g^{(2)}_{\mu \nu}) \ .
\end{align}
To find the correction proportional to $\alpha$, consider (\ref{g1}). In this term however the correction
\begin{align}
\Omega_{\mu}{}^\lambda \Omega_{\lambda \nu} + \frac{1}{d} P_{\mu \nu} \Omega_{\alpha \beta} \Omega^{\alpha \beta}
\end{align}
is traceless, so the metric determinant to second order is the same as in GR.

The null normal to horizon surface is defined as
\begin{align}
\ell^A = g^{AB} \partial_B (r-r_h(x)) \ ,
\end{align}
where in general the horizon location is a function of $x^\mu$. At second order, the result is
\begin{align}
\ell^{\mu~(2)} = g^{\mu r~(2)}|_{r=r_c-\frac{1}{p^2}}+ \frac{2}{p^3} P^{\mu \nu} \partial_\nu \ln p D \ln p - \frac{2}{p^3} P^{\mu \nu} \partial_\nu (D \ln p) \ .
\end{align}
Here we have used the first order inverse metric and
\begin{align}
r_h = r_c - \frac{1}{p^2} + \frac{2}{p^3} D \ln p + \cdots
\end{align}

Thus, any $\alpha$ corrections will be in $g^{\mu r (2)}$. However, this is given by
\begin{align}
g^{\mu r (2)} = p^{-1} P^{\mu \lambda} u^\sigma g^{(2)}_{\lambda \sigma} \ ,
\end{align}
which we saw above has no corrections. Thus, we conclude that $\ell^\mu_{(2)}$ in Einstein-Gauss-Bonnet is the same as in GR. Since $\ell^\mu_{(2)}u_\mu =0$, the first line of (\ref{secondentropy}) is just the area entropy current of GR. Thus,
\begin{align}
s^{\mu (2)}_{EGB} = s^{\mu (2)}_{GR} + s^{\mu (2)}_{GB} \ ,
\end{align}
where \cite{Compere:2012mt,Eling:2012ni}
\begin{align}
s^{\mu (2)}_{GR} &= 4\pi \left( \frac{1}{p^2} (K_{\alpha \beta} K^{\alpha \beta}) u^\mu + \frac{1}{2p^2} (\Omega_{\alpha \beta} \Omega^{\alpha \beta}) u^\mu \right) \nonumber \\ & +\frac{4\pi}{p^2} \left(2D^\mu_{\perp} \ln p D \ln p - 2 D^\mu_{\perp} D \ln p - P^{\mu \nu} \partial_\lambda \mathcal{K}^\lambda_\nu + (\mathcal{K}^\mu_\lambda + \Omega^{\mu}{}_\lambda) D^{\perp}_\lambda \ln p \right).
\end{align}
and
\begin{align}
s^{\mu (2)}_{GB} = 8\pi p^2 \mathcal{L}^{\mu \bar{r} \nu \bar{r}(2)}_{GB} u_\nu  + 16 \pi \alpha \mathcal{K}^{\mu \lambda}  D^{\perp}_\mu \ln p \ . \label{GBentropy}
\end{align}
Thus, we just need to compute the Gauss-Bonnet parts in (\ref{GBentropy}). For the details of the computation, see Appendix B. Ultimately we find that in terms of the fluid variables this has the form
\begin{align}
s^{\mu (2)}_{GB} =  8\pi \alpha (\mathcal{K}_{\alpha \beta} \mathcal{K}^{\alpha \beta}) u^\mu - 24\pi \alpha (\Omega_{\alpha \beta} \Omega^{\alpha \beta}) u^\mu - 16\pi \alpha P^\mu_\lambda \partial_\nu \mathcal{K}^{\nu \lambda} \ .
\label{GBentropyfinal}
\end{align}

Now let's consider the divergence of the total entropy current. At lowest order the divergence of the entropy current is just
\begin{align}
\partial_\mu s^{\mu} = 4\pi (\partial_\mu u^\mu)_{(2)} = \frac{8\pi}{p} \mathcal{K}^{\alpha \beta} \mathcal{K}_{\alpha \beta} \ ,
\end{align}
where in the last equality we have used the fluid equations. We can demonstrate the divergence is non-negative if we can show it still forms a perfect square at higher orders, i.e.
\begin{align}
\partial_\mu s^{\mu}_{EGB} = \frac{8\pi}{p} (\mathcal{K}_{\mu \nu} + S^{(2)}_{\mu \nu})^2 = \frac{8\pi}{p} \mathcal{K}^{\mu \nu}(\mathcal{K}_{\mu \nu} + 2 S^{(2)}_{\mu \nu}) + O(\epsilon^4) \ ,
\end{align}
where $S_{\mu \nu}$ is some second order tensor constructed from the fluid variables.

At third order, the divergence takes the form
\begin{align}
\partial_\mu s^{\mu (2)}_{EGB} = \partial_\mu s^{\mu (2)}_{GR} + 4\pi (\partial_\mu u^\mu)_{(3)} + \partial_\mu s^{\mu (2)}_{GB} \sim 2 \mathcal{K}^{\mu \nu} S^{(2)}_{\mu \nu} \ .
\end{align}
In the case of the divergence of the second order GR current, nothing changes from the previous calculations in (\cite{Compere:2012mt,Eling:2012ni}) and the divergence is of the perfect square form. However, the contribution from the divergence $\partial_\mu u^\mu$, which is determined by the third order fluid equation, $u^\mu \partial^\nu T_{\mu \nu} = 0$, does have $\alpha$ corrections. In the Gauss-Bonnet case, only the $c_{1,3}$ coefficients of the fluid stress receive $\alpha$ corrections (\ref{GBcoeff}) and the result is
\begin{align}
4\pi (\partial_\mu u^\mu)_{(3)} = -\frac{4\pi}{p^2}\mathcal{K}^{\alpha \beta} \left((2-4\alpha p^2) \mathcal{K}_{\alpha}^\lambda \mathcal{K}_{\beta \lambda} + 2 \mathcal{K}_{\alpha \beta} D \ln p - 12 \alpha p^2 \Omega_{\alpha}{}^\lambda \Omega_{\beta \lambda}\right) \ .
\end{align}
However, these $\alpha$ corrections are also of the required form and therefore will ultimately have no effect on whether the total current divergence is positive definite or not.

Finally, let's consider the divergence of the last purely Gauss-Bonnet term above. Taking the divergence and imposing the ideal fluid equations yields just
\begin{align}
\partial_\mu s^{\mu (2)}_{GB} = -16\pi \alpha \mathcal{K}^{\alpha \beta} \left(3 \Omega_{\alpha}{}^{\lambda} \Omega_{\beta \lambda} + \mathcal{K}_{\alpha}{}^\lambda \mathcal{K}_{\beta \lambda} \right) \ ,
\end{align}
which again has the necessary form for positive definiteness. Putting everything together, we find that
\begin{align}
s^{\mu}_{EGB} &= 4\pi \left(1+ \left(\frac{1}{p^2} + 2\alpha \right) (K_{\alpha \beta} K^{\alpha \beta}) + \left(\frac{1}{2p^2}-6\alpha \right)(\Omega_{\alpha \beta} \Omega^{\alpha \beta}) \right) u^\mu \nonumber \\ & +\frac{4\pi}{p^2} \left(2D^\mu_{\perp} \ln p D \ln p - 2 D^\mu_{\perp} D \ln p - (1+4\alpha p^2) P^{\mu \nu} \partial_\lambda \mathcal{K}^\lambda_\nu + (\mathcal{K}^\mu_\lambda + \Omega^{\mu}{}_\lambda) D_{\perp}^\lambda \ln p \right),
\end{align}
and
\begin{align}
\partial_\mu s^{\mu}_{EGB} &= \frac{8\pi}{p} \mathcal{K}^{\alpha \beta} \mathcal{K}_{\alpha \beta} \nonumber \\ & + \frac{4\pi}{p^2}\left(-8 \mathcal{K}^{\alpha \beta} D^{\perp}_\alpha D^{\perp}_\beta \ln p + 8 \mathcal{K}^{\alpha \beta} D^{\perp}_\alpha \ln p D^{\perp}_\beta \ln p - 4 D \ln p \mathcal{K}^{\alpha \beta} \mathcal{K}_{\alpha \beta} \right. \nonumber \\ & \left. - 8 \mathcal{K}^{\alpha \beta} \mathcal{K}^{\lambda}_{\alpha} \mathcal{K}_{\lambda \beta} - 8 \mathcal{K}^{\alpha \beta} \Omega_{\alpha}{}^{\lambda} \Omega_{\lambda \beta} \right).
\end{align}
Note that the $\alpha$ pieces coming from the field equations exactly cancel out with pieces arising from the explicit $\alpha$ parts of the current and we are left with just the GR result.

\section{Discussion}

In this work we considered the local horizon entropy current for higher-curvature gravitational theories, which in the non-stationary cases contains ambiguities related to the absence of a preferred timelike Killing vector. We argued that these ambiguities can be eliminated in general by choosing the vector that generates the subset of diffeomorphisms preserving a natural gauge condition on the bulk metric.
While this requirement determines the entropy current uniquely, it is not yet clear that it guarantees in general that it has a non-negative divergence.
As an example, we studied a dynamical, perturbed Rindler horizon in Einstein-Gauss-Bonnet gravity setting and computed the bulk dual solution to second order in fluid gradients. We showed that entropy current at second order has a manifestly non-negative divergence.

Notably, our condition (\ref{Killing1}) on $\ell^A$ has also appeared recently in \cite{Majhi:2012tf} outside the context of the fluid-gravity correspondence. Here the authors examine the approach to equilibrium horizon entropy using the Virasoro algebra, central charge, and the Cardy formula. One considers a subset of diffeomorphisms that in some sense respect the existence of a horizon in the bulk solution. Roughly, some of the full diffeomorphism symmetry is broken and some of the pure gauge degrees of freedom become physical. The authors argue that the subset consistent with (\ref{Killing1}) is preferred and apply it to a stationary Rindler metric. They show the vector $\ell^A$ obeys a Virasoro algebra with central extension and one can use the Cardy formula to count the asymptotic number of states. The result agrees with the Bekenstein-Hawking formula.
It would be interesting to understand better the nature of (\ref{Killing1}) and why it appears in different contexts.

In the fluid/gravity correspondence, condition (\ref{Killing1}) is necessary and sufficient to define an entropy current from the horizon data, which is compatible with the hydrodynamics dictated by the boundary stress-energy tensor. As noted above, it is an open problem to prove that this entropy current has in general a non-negative divergence. If true, it will prove in particular a Wald entropy increase theorem at least in the derivative expansion. A first step might be to consider the entropy current in Einstein-Gauss-Bonnet (or its Lovelock generalization), perhaps making use of the recent work in \cite{Chatterjee:2011wj,Kolekar:2012tq}, which proved the Wald entropy increases in quasi-stationary processes.

\section*{Acknowledgements}

We would like to thank Y. Neiman for a valuable discussion. C.E. thanks T. Nutma and A. Virmani for help with xAct \cite{xAct} which was used to check some of the calculations in this paper.
This work is supported in part by the Israeli Science Foundation center of excellence and by the German-Israeli Foundation (GIF).

\section*{Appendix A}

Here we present an alternative analysis of the ambiguities in the entropy current at first and second order in fluid gradients following \cite{Chapman:2012my}. In particular, here one does not a priori impose (\ref{Killing1}). At first order in gradients of the fluid variables, (\ref{entropycurrent}) contains the following set of unambiguous terms
\begin{align}
s^{(1) \mu}_{un-amb} = \frac{2\pi}{\kappa^{(0)}} \left(-2 \mathcal{L}^{\mu \bar{r} \nu \bar{r} (1)} A^{(0)}_\mu + 4 [\nabla_\nu \mathcal{L}^{\mu \bar{r} \nu \bar{r}}]^{(1)}\right) \ , \label{unamb}
\end{align}
and ambiguous terms
\begin{align}
s^{(1) \mu}_{amb} = -\frac{2\pi}{\kappa^{(0)}} \left(\mathcal{L}^{\mu \bar{r} \nu \bar{r} (0)} A^{(1)}_\nu + \kappa^{(1)} u^\mu \right) \ .
\end{align}
The ``ambiguous" terms are associated with how one defines the surface gravity and radial derivatives of the null normal away from equilibrium.  Note, that $[\nabla_\nu \mathcal{L}^{\mu \bar{r} \nu \bar{r}}]^{(0)}= 0$ for any $\mathcal{L}^{ABCD}$. Using the fact that generically
\begin{align}
\mathcal{L}^{\mu \bar{r} \nu \bar{r} (0)} = \frac{1}{2} u^\mu u^\nu \\
A^{(0)}_\mu =  -2 \kappa^{(0)} u_\mu \ ,
\end{align}
we find that this ambiguity is eliminated if
\begin{align}
u^\mu A^{(1)}_\mu = 2 \kappa^{(1)} \ . \label{weakKilling1}
\end{align}
One can show that this is equivalent to (\ref{weakKilling0}) at first order evaluated on the horizon. If we impose this condition, then to first order there is no ambiguity in the entropy current for any higher curvature theory of gravity. The unambiguous part of the current was computed in \cite{Chapman:2012my} in the case of a charged black brane background in Einstein-Maxwell theory with a mixed gauge-gravitational Chern-Simons term. The resulting entropy current agrees with the field theory calculation and has a non-negative divergence, indicating that this current may be the correct one.

At second order in gradients,
\begin{align}
s^{\mu (2)}_{amb} = \frac{2\pi \sqrt{-g}}{\kappa^{(0)}} \left(-2 \mathcal{L}^{\mu \bar{r} \nu \bar{r} (1)} A^{(1)}_\nu + \frac{1}{2} u^\mu (u^\nu A^{(2)}_\nu) - \kappa^{(2)} u^\mu \right) \ . \label{2ndamb}
\end{align}
We can express $\mathcal{L}^{\mu \bar{r} \nu \bar{r} (1)}$ in terms of the fluid variables as follows
\begin{align}
\mathcal{L}^{\mu \bar{r} \nu \bar{r} (1)} =  \alpha^{(1)} u^\mu u^\nu + \beta^{(1)} P^{\mu \nu} + 2  u^{(\mu} Y^{\nu)}_{(1)} + \gamma \sigma^{\mu \nu} \ , \label{genLfirst}
\end{align}
where $\alpha^{(1)}$ and $\beta^{(1)}$ are first order scalars constructed out of the fluid variables, $Y^{(1)}_\mu$ is a first order vector, and $\sigma_{\mu \nu} = P^\lambda_\mu P^\sigma_\nu \partial_{(\lambda} u_{\sigma)}-\frac{1}{d} P_{\mu \nu} \partial_\lambda u^\lambda$. In general, these ambiguous terms have a somewhat complicated form relating the components of $A_\mu$ to $\kappa$ and it appears that we can say little without actually computing the perturbed solution in a particular theory to second order in gradients.

In the Rindler case above, the ambiguities reduce to just
\begin{align}
s^{\mu (2)}_{amb} = 4\pi p (u^\nu A^{(2)}_\nu) u^\mu - 8\pi \kappa^{(2)} u^\mu - 4\pi p \mathcal{L}^{\mu \bar{r} \nu \bar{r}(1) } A^{(1)}_\nu,   \label{amb1st1st}
\end{align}
via (\ref{firstorderL}). Essentially in (\ref{genLfirst}) only the $\gamma \sigma^{\mu \nu}$ term is non-zero. The first two terms proportional to $u^\mu$ above vanish if we impose the weak Killing condition at second order $u^\mu A^{(2)}_\mu = 2 p \kappa^{(2)}$. To compute the remaining term, we can use the previous results in (\ref{firstorderL}) and (\ref{zerothdata}) to show that
\begin{align}
s^{\mu (2)}_{amb} = -8\pi \alpha \mathcal{K}^{\mu \nu} A^{(1)}_\nu. \label{Rindleramb}
\end{align}
Thus, the remaining ambiguity is in the other set of $d$ degrees of freedom contained in $P^\nu_\mu A^{(1)}_\nu$.

If we repeat the calculation of the entropy current with (\ref{Rindleramb}), we find
\begin{align}
\partial_\mu s^{\mu (2)}_{GB} &= -16\pi \alpha \mathcal{K}^{\alpha \beta} \left( D^{\perp}_{\alpha} D^{\perp}_\beta \ln p + 3 \Omega_{\alpha}{}^{\lambda} \Omega_{\beta \lambda} + \mathcal{K}_{\alpha}{}^\lambda \mathcal{K}_{\beta \lambda} - \frac{1}{2} D^{\perp}_\alpha A^{(1)}_\beta\right) \nonumber \\ &  - 16 \pi \alpha (\partial_\mu \mathcal{K}^{\mu \lambda}) D^{\perp}_\lambda \ln p - 8\pi \alpha (\partial_\mu \mathcal{K}^{\mu \lambda}) A^{(1)}_\lambda.
\end{align}
Therefore we see that the divergence will not be of the perfect square form in general unless we require exactly the last condition in (\ref{firstdata}), that $P^\lambda_\mu A^{(1)}_\lambda = - 2 D^{\perp}_\mu \ln p$. Using this information, we can work backwards and compute the form of $\ell^A$. The result is
\begin{align}
\ell_A dX^A = 2 p \kappa \bar{r} u_\mu dx^\mu + 2 \bar{r} (D^{\perp}_\mu \ln p) dx^\mu  + d\bar{r} + \bar{r} f^{(1)} d\bar{r}, \label{ellower}
\end{align}
We have parameterized the extra degree of freedom in $\ell^{(1)}_{\bar{r}} = \bar{r} f^{(1)}$. The conditions for positivity of the divergence of the entropy current seem to be just $(d+1)$ conditions on the $(d+3)$ degrees of freedom in $\kappa^{(1)}$ and $\ell^{(1)}_A$. However, we can actually calculate $\kappa^{(1)}$ using the formula
\begin{align}
\ell^\mu \nabla_\mu \ell_r = \kappa \ell_r  \rightarrow -[\ell^\mu \Gamma^{\bar{r}}{}_{\mu \bar{r}}]^{(1)}|_{\bar{r}=0} = \kappa^{(1)}. \label{kappaeqn}
\end{align}
Raising (\ref{ellower}) to a vector yields
\begin{align}
\ell^A \frac{\partial}{\partial X^A} = \frac{1}{p} \left(1+ f^{(1)} \bar{r}\right) u^\mu \frac{\partial}{\partial x^\mu} + \bar{r} \left(\bar{r} f^{(1)} + 2 \kappa^{(1)} - \frac{2}{p} D \ln p \right) \frac{\partial}{\partial \bar{r}}
\end{align}
Using this result and the first order metric, (\ref{kappaeqn}) yields $\kappa^{(1)} = p^{-1} D \ln p$ and thus
\begin{align}
\ell^A \frac{\partial}{\partial X^A} = \frac{1}{p} \left(1+ f^{(1)} \bar{r}\right) u^\mu \frac{\partial}{\partial x^\mu} + \bar{r}^2 f^{(1)} \frac{\partial}{\partial \bar{r}}.
\end{align}
Thus, up to the freedom in $f^{(1)}$ this result agrees with (\ref{ellvector}). Setting $f^{(1)}=0$ is exactly what we need for (\ref{Killing1}) to hold.

\section*{Appendix B}

In this appendix we show the detailed calculation of the Gauss-Bonnet part of the entropy current in (\ref{GBentropy}).  Using (\ref{LGB}), we find
\begin{align}
\mathcal{L}^{\mu \bar{r} \nu \bar{r}(2)}_{GB} u_\nu &= 2\alpha( R^{\mu \bar{r} \nu \bar{r}(2)} u_\nu - p^{-1} R^{\mu \bar{r} (2)} + p^{-1} (R^{\nu \bar{r} (2)} u_\nu) u^\mu + \frac{1}{2} p^{-2} R^{(2)} u^\mu) \label{secondorder}
\end{align}
Computing the Ricci tensor components and Ricci scalar is easier because we are working on-shell, which implies they are zero up to second order. Thus the transformation between the $(r,x^\mu)$ and $(\bar{r},x^\mu)$ coordinates
\begin{align}
\frac{\partial{\bar{r}}}{\partial r} =& 1 \nonumber \\
\frac{\partial{\bar{r}}}{\partial x^\mu} =& - \partial_\mu r_h \label{transform}
\end{align}
is trivial at second order and we can work in the original set of coordinates. The equations of motion (\ref{eom}) imply
\begin{align}
R^{(2)} &= \frac{4 \alpha}{d} \hat{H}^{(2) C}_C \\
R^{\mu r (2)} &= \frac{2\alpha}{d p} u^\mu \hat{H}^{(2) C}_C - \frac{2 \alpha}{p} P^{\mu \lambda} u^\sigma \hat{H}^{(2)}_{\lambda \sigma} \label{eomsecondorder}
\end{align}
The result is
\begin{align}
R^{\mu \bar{r} (2)} &= 3\alpha p \left(\frac{d-2}{d}\right) (\Omega_{\alpha \beta} \Omega^{\alpha \beta}) u^\mu \\
R^{(2)} &= 6 \alpha p^2 \left(\frac{d-2}{d}\right) \Omega_{\alpha \beta} \Omega^{\alpha \beta}.
\end{align}
Note that these corrections already depend on $\alpha$, which means they will be, in principle, of $O(\alpha^2)$ in the entropy current formula. Inserting these results into (\ref{GBentropy}) we find
\begin{align}
s^{\mu (2)}_{GB} &= 16\pi p^2 \alpha R^{\mu \bar{r} \nu \bar{r} (2)} u_\nu - 48 \alpha^2 p^2 \left(\frac{d-2}{d}\right) (\Omega_{\alpha \beta} \Omega^{\alpha \beta}) u^\mu
- 8\pi \alpha \mathcal{K}^{\mu \lambda} A^{(1)}_\lambda. \label{GBentropy2}
\end{align}

Next, one can relate $R^{\mu \bar{r} \nu \bar{r} (2)}$ to the unbarred $r$ components using the coordinate transformation (\ref{transform})
\begin{align}
R^{\mu \bar{r} \nu \bar{r}} = \frac{\partial \bar{r}}{\partial X^A} \frac{\partial \bar{r}}{\partial X^B} R^{\mu A \nu B},
\end{align}
which yields
\begin{align}
R^{\mu \bar{r} \nu \bar{r} (2)} u_\nu = R^{\mu r \nu r (2)} u_\nu - 2 p^{-2} \partial_\lambda \ln p (R^{\mu \lambda \nu r (1)} u_\nu + R^{\nu \lambda \mu r (1)} u_\nu). \label{Rtransform}
\end{align}
Using
\begin{align}
R^{r \mu \nu \lambda }_{(1)} =  u^\mu \Omega^{\nu \lambda} + \frac{1}{2} u^\nu \Omega^{\mu \lambda} - \frac{1}{2} u^\lambda \Omega^{\mu \nu},
\end{align}
we find
\begin{align}
R^{\mu \bar{r} \nu \bar{r} (2)} u_\nu = R^{\mu r \nu r (2)} u_\nu - 3 p^{-2} \Omega^{\mu \lambda} D^{\perp}_\lambda \ln p
\end{align}
As a result, (\ref{GBentropy2}) becomes
\begin{align}
s^{\mu (2)}_{GB} &= 16 \pi \alpha p^2 R^{\mu r \nu r (2)} u_\nu - 48 \alpha^2 p^2 \left(\frac{d-2}{d}\right) (\Omega_{\alpha \beta} \Omega^{\alpha \beta}) u^\mu  - 48 \pi \alpha \Omega^{\mu \lambda} D^{\perp}_\lambda \ln p - 8\pi \alpha \mathcal{K}^{\mu\lambda} A^{(1)}_\lambda .
\end{align}
Finally, by direct computation using our metric solution, we find
\begin{align}
{R}^{\mu r \nu r}_{(2)} u_\nu &= \frac{1}{p^2} \left(-P^{\mu \lambda} \partial_\rho \mathcal{K}^\rho_\lambda -  \mathcal{K}^{\mu \lambda} D^{\perp}_\lambda \ln p + 3 \Omega^{\mu \lambda} D^{\perp}_\lambda \ln p \nonumber \right. \\ & \left. + \frac{1}{2} (K_{\alpha \beta} K^{\alpha \beta}) u^\mu + \frac{3}{2} [2\alpha p^2 (\frac{d-2}{d})-1] (\Omega_{\alpha \beta} \Omega^{\alpha \beta}) u^\mu   \right).
\end{align}
Putting all these components together yields (\ref{GBentropyfinal}) and the $O(\alpha^2)$ pieces ultimately cancel out.

\end{document}